\documentclass{epl}
\usepackage{amsmath}

\title{Inverse proximity effect in superconductors near ferromagnetic material}
\shorttitle{Inverse proximity effect etc.}
\author{M. A. Sillanp\"a\"a\inst{1}\thanks{E-mail: \email{Mika.Sillanpaa@iki.fi}}
\and T. T. Heikkil\"a\inst{2,3} \and R. K. Lindell\inst{1} \and P. J. Hakonen\inst{1}}
\shortauthor{M. A. Sillanp\"a\"a \etal}
\institute{
 \inst{1} Low Temperature Laboratory, Helsinki University of Technology - P.O.Box 2200, FIN-02015 HUT, Finland\\
 \inst{2} Materials Physics Laboratory, Helsinki University of Technology - P.O.Box 2200, FIN-02015 HUT, Finland\\
 \inst{3} Institut f\"ur Theoretische Festk\"orperphysik, Universit\"at Karlsruhe - D-76128 Karlsruhe, Germany
}
\pacs{74.50.+r}{Proximity effects, weak links, tunneling phenomena, and Josephson effects}

\begin{document}

\maketitle

\begin{abstract}
We study the electronic density of states in a mesoscopic superconductor near
a transparent interface with a ferromagnetic metal.
In our tunnel spectroscopy experiment,
a substantial density of states is observed at sub-gap energies close to a ferromagnet.
We compare our data with detailed calculations based on the Usadel equation,
where the effect of the ferromagnet is treated as an effective boundary condition.
We achieve an excellent agreement with theory when non-ideal
quality of the interface is taken into account.
\end{abstract}

Proximity effect (PE) is a phenomenon where the order parameter of a
material in an ordered phase leaks to another material which has no
such order. A number of experiments have investigated the PE in
superconducting - normal (S-N) mesoscopic structures \cite{nprox}. Here
a finite pairing amplitude is formed in the N side, with the decay
length  $\xi_{N} = \sqrt{\hbar D/ 2 \pi k_{B}T}$ set by the dephasing
of correlations between electrons and Andreev-reflected holes
having slightly different momenta. At typical experimental conditions at
low temperatures, the effect can survive up to large distances of
several hundred nanometers. The experiments in non-magnetic metals
have been generally well understood within the framework of the
quasiclassical theory of inhomogeneous superconductivity
\cite{belzig}. Another example of a proximity effect occurs in systems
composed of ferromagnetic (F) and nonmagnetic materials, where the
spin polarization leaks from F \cite{spin}, but its decay length is
typically very small. Going further, one may also consider the
combination of two differently ordered materials, such as
superconductors and ferromagnets. 

In this Letter, we study the mutual proximity effects of
superconducting and ferromagnetic materials by measuring the local
density of states in the superconducting side of the SF interface
formed by superconducting Al and ferromagnetic Ni. Since the
ferromagnetic and singlet superconducting order parameters try to
exclude each other, the direct proximity effect is expected to get 
suppressed \cite{buzdin} such that the pairing amplitude diffusing
into a ferromagnet should decay over microscopic distances $\xi_{F} =
\sqrt{\hbar D/ 2 \pi k_{B} T_{\rm Curie}}$ due to the considerable
difference in the momenta of spin-up and spin-down quasiparticles.  A
spectroscopy experiment in a low-$T_{\rm Curie}$ ferromagnet
\cite{fdosmeas} agreed with the theoretical picture of this
short-range effect. However, some recent experiments
\cite{petra_ferro1,giroud, petra2, petra3} appear to indicate the
existence of a long-range (i.e. a penetration depth comparable to
$\xi_{N}$) proximity effect into a ferromagnet and the situation has
thus remained elusive. Here we report the first measurement on an 
inverse phenomenon, the modification of the BCS density of states in
mesoscopic superconducting strips of Al under the influence of the 
proximity effect of a classical ferromagnet (Ni). This is rather a
result from the expected suppression of the proximity effect: since
the pairing amplitude vanishes at the SF interface, its value within a
short distance from the interface is also much smaller than in the
corresponding case of a finite proximity effect into a normal metal. 

We use tunnel spectroscopy with tunnel probes at fixed positions to
measure the differential conductance on the S side of
the interface. At zero temperature, and for a vanishing charging
energy of the tunnel contact, the differential
conductance $G(E)$ with an N metal probe is proportional to the density of
states (DOS). This way we can probe how the DOS in a superconductor is
affected by the proximity effect.
The DOS on the N side of an SN proximity system has been measured with
fixed tunnel probes by Gueron
\textit{et al.} \cite{gueron}, who observed a dip in the N metal DOS
up to one $\mu$m from the interface, with a quantitative agreement
to theory.
Several STM experiments capable of spatial mapping of the DOS have been carried
out in short-coherence length layered superconductors \cite{stm_93, stm_96, stm_99},
and recently in Nb-Au structure \cite{grenoble}.

\begin{figure}[bp]
\onefigure[width=0.6\linewidth]{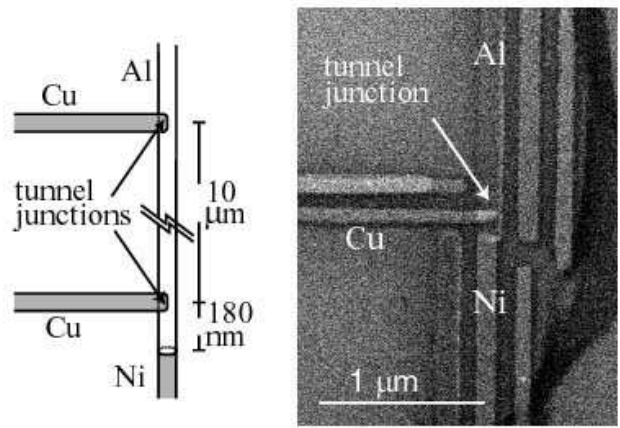}
\caption{Schematics of the sample, and SEM micrograph showing the region
near the SF-interface. The pattern was chosen so that the additional replicas due
to shadow evaporation did not interfere with the desired pattern. Width of
the lines was 100-150 nm.}
\label{sample}
\end{figure}

We used a sample structure shown in Fig.\ \ref{sample}.
In the sample, we have two tunnel
spectroscopy probes: one at a distance of approximately 180 nm from the
transparent SF interface (called the interface
junction), and one 10 $\mu$m from the interface (called the bulk junction).
The latter tunnel junction is assumed to be located at a site
with properties of bulk Al.
The sample was fabricated on oxidized Si wafers by
electron beam lithography and shadow evaporation through PMMA/MMA
copolymer mask. We first evaporated 20 nm of Al.
To make the clean SF interface, we evaporated a 25 nm layer of Ni at a different
angle. The depositions of
the first two layers were done in a single UHV cycle at pressures below
$5 \cdot 10^{-9}$ mBar, avoiding any delay between the successive layers
to make a transparent interface. The Al wires were then oxidized in 0.3
mBar of 0$_{2}$ for 5 minutes in order to make the opaque barriers for tunnel
spectroscopy. Lastly, the tunnel probes were made by depositing 40 nm of Cu at
a third angle. The process yielded resistances of approximately 1.5 M$\Omega$
for both tunnel junctions.

For electric transport measurements, the sample was cooled down to
approximately 100 mK in a plastic dilution refrigerator.
Conductance measurements for the tunnel probes were performed through
carefully filtered coaxial leads. To measure
differential conductance, we used lock-in techniques with 20-30 $\mu$V ac
excitation added to a constant dc voltage bias
applied to the tunnel probe to be measured.

\begin{figure}[bp]
\onefigure[width=0.6\linewidth]{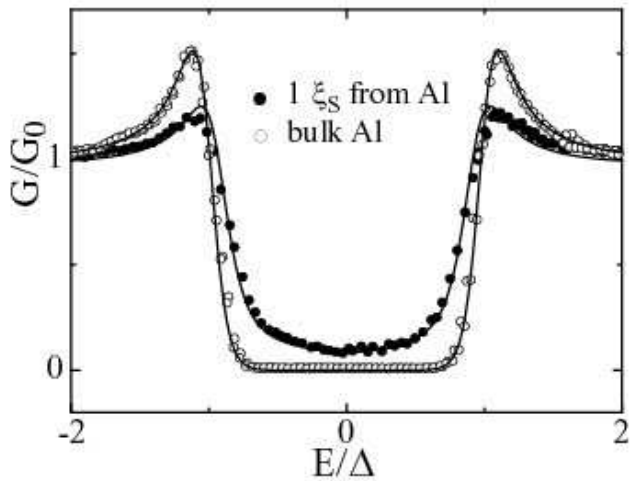}
\caption{Normalized differential conductance $G/G_{0}$ of the SIN junction,
when properties of S are affected by the proximity of a ferromagnet. The circles are the
experimental data at a distance of 180 nm ($1.2 \; \xi_{S}$) from the
interface, and that for the bulk Al, shown
for reference (measured at T = 100 mK). The data are normalized to the conductance value $G_{0}$
measured at 1 mV. The solid curves are the best fits to theory,
with $\alpha = 7.5$ (implying $R_{\rm env} = 870 \; \Omega)$,
$E_{C} / \Delta = 1.5$, $r_{b} = 3$.
The gap of Al was $\Delta = 0.22$ mV.}
\label{data}
\end{figure}

The experimental results are shown in Fig.\ \ref{data}.
For the bulk junction, we obtained a BCS-like differential conductance, although
the BCS peaks were somewhat lower than what would be expected
simply by thermal broadening. The magnitude of the peaks was accurately
accounted for only when we took into account the
Coulomb charging effects (see below). For the interface junction, the peaks
were lower, with a clear sub-gap conductance.

In the normal state, superconductivity being suppressed by perpendicular magnetic field,
we observed a conductance dip around small bias due to Coulomb charging effects.
Both tunnel probes showed a dip of nearly 40 \% due to a resistive
environment created by the relatively resistive nickel wires.

Diffusion constants for the wires or the interface resistance were not directly
measurable in the sample. We used test samples, made in a similar process,
to measure resistivities for thin wires of Al, Ni and Cu,
with the results $\rho_{Al} = 1.7 \ \mu \Omega$cm, $\rho_{Cu} = 2.0 \ \mu \Omega$cm
and $\rho_{Ni} = 32 \ \mu \Omega$cm.

The density of states near a SF interface can be calculated from
the Usadel equation \cite{usadel} for the quasiclassical
Green's functions in the diffusive limit. These functions can be
parametrized using the $\theta$-parametrization \cite{belzig}. In
the absence of supercurrents the retarded Green's function is
$\hat{G}^R=\cosh(\theta)\hat{\tau}_3+i\sinh(\theta)\hat{\tau}_2$,
where $\hat{\tau}_i$ are the Pauli matrices in Nambu space. In the
case of translational invariance in the transverse directions, the
Usadel equation then takes the form 
\begin{equation}
\hbar D \partial_x^2 \theta = -2iE \sinh(\theta(E,x))+2i\Delta(x)
\cosh(\theta(E,x)).
\label{eq:usadel}
\end{equation}
Here $D$ is the diffusion constant and $\Delta(x)$ is the
superconducting pair potential, calculated self-consistently
from 
\begin{equation}
\Delta(x)=\lambda N_0 2\pi k_{B} T \sum_{\omega_n} \sinh{\theta(\omega_n,x)},
\label{eq:selfconsistency}
\end{equation}
where $N_0$ is the density of states in the normal state, and the sum
goes over the discrete Matsubara frequencies $\omega_n=(2n+1)\pi k_B T$
and is cut off at the Debye energy. The attractive interaction between
the electrons is characterized by the coupling parameter $\lambda$,
which is a nonzero constant for superconducting materials and zero
otherwise. The parameter $\theta(\omega_n,x)$ is calculated from Eq.\
(\ref{eq:usadel}), where the energy $E$ is replaced by $i \omega_n$.

To account for a possible non-ideality of the interface between the two materials
in the experiment, Eqs.\ (\ref{eq:usadel},\ref{eq:selfconsistency}) have to be
complemented by a boundary condition recently derived by Nazarov
\cite{nazarov}, relating the Green's function in the left side of the
interface ($\hat{G}_1=\hat{G}(x=0^-)$) to the function in the right
side ($\hat{G}_2=\hat{G}(x=0^+)$), 
\begin{equation}
\sigma_1 \hat{G}_1 \partial_x \hat{G}(0^-) = \sigma_2 \hat{G}_2 \partial_x
\hat{G}(0^+) =
\frac{e^2}{h} \sum_n
\frac{ 2 {\cal T}_n \left[\hat{G}_2,\hat{G}_1\right]}{4+{\cal T}_n\left(\left\{\hat{G}_2,\hat{G}_1\right\}-2\right)}.
\label{eq:nazarovbc}
\end{equation}
In a junction with many transmission channels, the distribution of the
transmission eigenvalues ${\cal T}_n$ can be obtained from random matrix
theory \cite{belzig00}. In the case of a dirty interface it is given
by \cite{schep}
\begin{equation}
\rho({\cal T})=\frac{\hbar}{\rho_B e^2}\frac{1}{{\cal T}^{3/2}\sqrt{1-{\cal T}}},
\label{eq:transmissiondistribution}
\end{equation}
where $\rho_B$ is the normal-state resistance of the interface per
unit area. Integrating Eq.\ (\ref{eq:nazarovbc}) over the distribution
of eigenvalues and over the cross sections of the wires,
we finally get the desired boundary condition
\begin{equation}
\partial_x\theta(0^+)=\frac{R_\xi}{R_{l\xi}}\partial_x\theta(0^-)=\frac{\sqrt{2}}{r_b}
\frac{\sinh(\theta(0^+)-\theta(0^-))}{\sqrt{1+\cosh(\theta(0^+)-\theta(0^-))}},
\label{eq:bc}
\end{equation}
where $r_b=R_B/R_\xi$ is the ratio between the
normal-state resistance $R_B=\rho_B A_B$ of the interface with the cross
section $A_B$, and the normal-state resistance $R_\xi=\xi_S/\sigma_N A$ of a piece of
the right-hand-side (S) wire with length $\xi_S$, conductivity $\sigma_N$
and cross section $A$. The
corresponding resistance of the wire in the left (F) side of the interface is
denoted by $R_{l \xi}$.

\begin{figure}[bp]
\onefigure[width=0.9\linewidth]{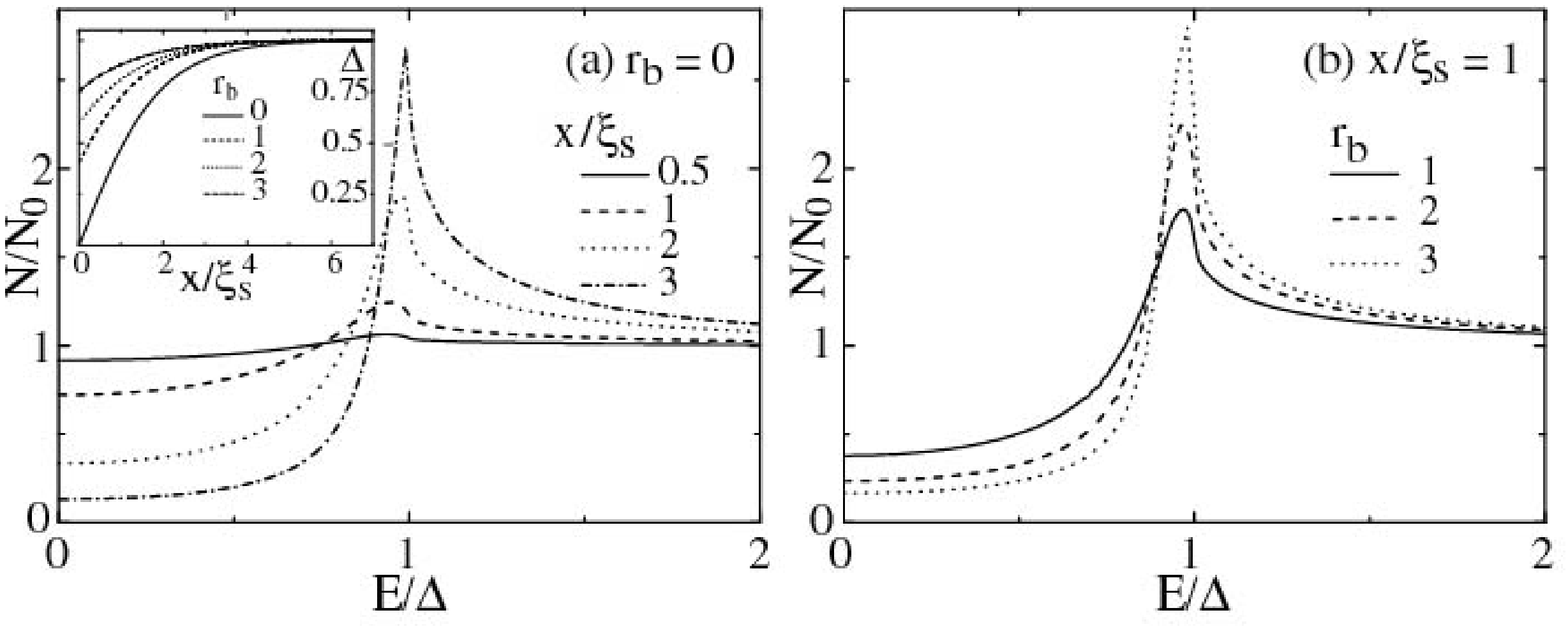}
\caption{DOS of SF structures calculated on the S side, (a) with an
ideal interface, $r_b = 0$, at distances $0.5 \; \xi_{S} ... 3 \; \xi_{S}$ from the interface;
(b) at a fixed distance $x=\xi_{S}$ with varying interface parameters $r_b = 1...3$.
In the calculations, the
S order parameter was assumed not to penetrate F, i.e. DOS in F was taken as constant $N_{0}$.
The inset shows the self-consistent pair potential.}
\label{f_dos}
\end{figure}

From the solution $\theta(E,x)$ to Eqs.\
(\ref{eq:usadel},\ref{eq:selfconsistency},\ref{eq:bc}), we obtain the
local density of states through 
\begin{equation}
N(E,x)=N_0 {\rm Re}\{\cosh(\theta(E,x))\}.
\label{eq:dos}
\end{equation}
In the ferromagnet, we assume the proximity effect to
vanish within a few nm from the interface, and therefore set
$\theta(x=0^-)=0$. In Fig.\ \ref{f_dos} we plot the DOS at varying values of
the interface parameter $r_{b}$
and at varying distances to the interface. Since coupling to the interface
decreases with increasing $r_b$, DOS is very sensitive to the interface
transparency.

To calculate differential conductance of a tunnel probe, we have to take into account
Coulomb charging effects due to the smallness of the tunnel junction capacitance.
The charging effects depend on the impedance of the electromagnetic environment
in terms of the function $P(E)$ which is the probability of exchange of energy between the
tunneling electron and the environment. Forward tunneling rate through a single
junction having the tunnel resistance $R_{T}$ is given by \cite{ingnaz}

\begin{equation}
\Gamma(V) = \frac{1}{e^{2} R_{T}} \int ^{+\infty} _{-\infty} \mathrm{d}E \mathrm{d}E'
\frac{N_{0} N_{S}(E' + eV)}{N_{0}^{2}}
f(E) [1-f(E'+eV)]P(E-E'),
\label{eq:uelfinal}
\end{equation}
where $N_{0}$ is DOS of the tunnel probe, taken as constant, and $N_{S}$
is the proximity-affected DOS on the S side. The function $P(E)$ is calculated
from an integral equation assuming a purely resistive environment
\cite{ingold_peint}.

For both tunnel probes, the parameters $\alpha =
R_{Q}/R_{\rm env}$ and the charging energy of
tunnel junction capacitor $E_{C}$ were determined as the values that gave the best fit
to the normal-state conductance. Here, $R_{Q} = h/4e^{2} \simeq 6.5$ k$\Omega$
is the superconducting resistance quantum, and $R_{\rm env}$ is the resistance
of the electromagnetic environment seen by the tunnel junction. Normal-state
data of both junctions were fitted with the same values.
These values were then used to fit $\Delta$
in bulk Al data, and we obtained a faultless fit with $\Delta \simeq 0.22$ mV and with
the temperature $T = 100$ mK that was recorded in the experiment.

The distance between the interface and the first probe
had some ambiguity of $\pm 50$ nm because
the tunnel probes had a finite width of 80 nm, and because
the two metal films at the interface overlapped each other by approximately 50 nm.
Average of the DOS over these distances
was practically equal to the DOS in the middle of the tunnel probe, and we
used the average distance (this distance is about $1.2 \; \xi_{S}$) in the analysis.

The normal-state and bulk-junction measurements yielded all the other
important parameters of the system except the interface parameter $r_b$.
For example, the electronic temperature in both
bulk and interface junctions can be assumed the same due to the
similarity of the probe resistances.
Typical resistances reported for SF interfaces similar to our sample
have been in the range $20 \Omega -30 \Omega$ \cite{petra3}.
For our sample made in UHV conditions,
we expect $R_B \simeq 10 \ \Omega$. Since $R_{\xi} \approx 2 \; \Omega$,
we would have $r_b \simeq 5$.

With a reasonable value of the interface parameter $r_{b} = 3$,
we get an excellent fit with the theory, as seen in Fig.\ \ref{data}.
Note that this fit was obtained without taking
into account such effects as the magnetic field due to the
ferromagnet, penetration of the spin polarization in the
superconducting side, or the emergence of a triplet proximity effect
\cite{bergeret}.

The effect of a stray magnetic field from the ferromagnet
can be examined as follows. We expect the
size of magnetic domains in the film to equal the thickness of the film,
$d = 25$ nm. Since the dipolar field from a single domain decays with distance $r$
as $(d / r)^3$, at relevant distances $r \simeq 200$ nm, the saturation field
of Ni of 0.6 T has dropped down to 1 mT. Compared to the much larger
critical field of the films (70 mT), this is unlikely to explain the enhancement of
DOS at low energies. To gain further confidence that the observed DOS was
not affected by the stray magnetic field, we measured the conductance at
various externally applied field strengths. At fields up to $\sim 20$ mT,
the zero-bias conductance stayed practically constant while the width
of the gap decreased. This behavior was similar in both
the interface junction and the junction at bulk Al. Thus, stray fields
even significantly larger than the estimated would not be likely to affect the zero-bias
DOS, but would only reduce the gap. The case is strengthened by the
fact that the gap was not reduced more than expected from
the Usadel theory. These arguments also favor the idea that
the domains were indeed of the estimated size.

In conclusion, we have measured the inverse proximity effect in a
mesoscopic superconductor in contact with Ni, and compared the data
with a self-consistent calculation based on the Usadel equation where
we assume Ni to suppress the PE in a microscopic scale. We obtain a
good agreement for the electronic density of states close to the
interface only by taking into account the effect of the non-ideal
interface.

\acknowledgments
The authors would like to thank \textsc{M. Giroud} and \textsc{W. Belzig} for useful ideas
and comments. This research was supported in part by Emil Aaltonen foundation,
and by the Human Capital and Mobility Program \textsc{Ulti} of the European Union.

\end{document}